\documentclass[aps,twocolumn,showpacs]{revtex4}
\usepackage{graphics}

\begin{document}
\title{Spectral and polarization dependencies of luminescence by hot
carriers in graphene}
\author{F.T. Vasko}
\email{ftvasko@yahoo.com}
\affiliation{Institute of Semiconductor Physics, NAS of Ukraine, Pr. Nauki 41,
Kiev, 03028, Ukraine}
\author{O.G. Balev}
\affiliation{Departamento de Fisica, Universidade Federal do Amazonas,
Manaus, 69077-000, Brazil}
\date{\today}

\begin{abstract}
The luminescence caused by the interband transitions of hot carriers in graphene
is considered theoretically. The dependencies of emission in mid- and near-IR
spectral
regions versus energy and concentration of hot carriers are analyzed;
they are determined both by an applied electric field and a gate voltage. The
polarization dependency is determined by the angle between the propagation direction
and the normal to the graphene sheet. The characteristics of radiation from
large-scale-area samples of epitaxial graphene and from microstructures of
exfoliated graphene are considered. The averaged over angles efficiency of emission
is also presented.
\end{abstract}


\maketitle

\section{Introduction}
Electro- and photoluminescence of the bulk semiconductor materials and structures
have been used for more than fifty years both in the characterization these materials
and in the operation of light-emitting devices. \cite{1} Emission of radiation in
the far- and mid-IR spectral regions caused by nonequilibrium charge carriers
also have been studied for two-dimensional systems, see reviews in Ref. \cite{2},
and for the transitions between the subbands of heavy and light holes in $p$-Ge. \cite{3}
In graphene processes of emission of radiation  are actual due to the effective
heating of
carriers by dc electric field \cite{4,5,6} and due to the the efficient  interband transitions excited by photons. \cite{7} In this work we analyze spectral
and polarization dependences of emission for interband transitions (on frequencies
exceeding the frequency of relaxation) induced by hot carriers in a bipolar graphene.
The characteristics of radiation are considered for two cases of emission: (a)
from large-scale-area samples of epitaxial graphene and (b) from microstructures
of exfoliated graphene, see Figs. 1a and 1b, respectively. The analysis is based on
the quasiclassical kinetic equation for 3D photons where the interaction with 2D
carriers is described by the boundary condition at graphene sheet (see Ch. 4 in
\cite{8} and a similar approach for the case of acoustic phonon emission \cite{9}).

\begin{figure}[ht]
\begin{center}
\includegraphics{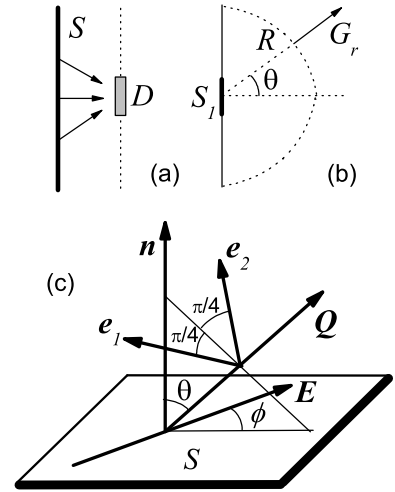}
\end{center}
\addvspace{-3.5 cm}
\caption{Geometry of luminescence by hot carriers from graphene sheet of: (a)
a large-scale-area sample ($S$), with detector ($D$) placed in the near-zone;
(b) a finite-size sample ($S_1$), where $G_r$ is the energy flow in the
far-zone. (c) Polarization characteristics of radiation. Here $\textbf{n}$ is the
normal to 2D plane, $\textbf{e}_{1,2}$ are the unit vectors of polarization, and
$\textbf{Q}$ is the wave vector.
Angles $\theta$ and $\phi$ define propagation
direction and in-plane orientation, respectively. }
\end{figure}

Spectral dependencies of emission are determined by the character of distributions
of nonequilibrium electrons and holes, wherein a heating electric field modifies
not only an effective temperature of carriers but also electron and hole concentrations.
These nonequilibrium characteristics are dependent on a lattice temperature, $T$,
a strength of applied electric field, $\textbf{E}$, and a gate voltage $V_{g}$. Due
to direct interband transitions and a linear character of the energy spectrum of
graphene, a radiation with frequency $\omega$ is emitted for transitions between
electron ($e$) and hole ($h$) states with the momentum $p_{\omega}=\hbar\omega
/2v_{W}$, where $v_{W}\approx 10^{8}$cm/s is the characteristic velocity of the
Weyl-Wallace model (degenerated both on spin and valley quantum numbers). \cite{10}
It appears that spectral dependence of radiation is proportional to the product of
$e$- and $h$-distribution functions, $f_{e p_{\omega}}f_{h p_{\omega}}$, and the
maximum energy of emitted photons is determined by both the effective temperature
of carriers, $T_{c}$, and the electron and hole concentrations. The spectral behavior
of radiation is strongly dependent on the densities of carriers, which in turn are essentially dependent on $V_{g}$.

The polarization dependence of emission results from the chiral character of the
neutrino-like states, \cite{11} so that the matrix elements of interband transition
are dependent on orientation of 2D momentum, $\textbf{p}$. \cite{5}  Due to this,
a strong dependence of polarization from the angle $\theta$, between the direction
of propagation of radiation and the normal vector to the graphene sheet (see Fig. 1),
is obtained. In addition, a weak the current-induced anisotropy of distributions,
$\propto (eE\tau_{m}/\bar{p})^{2}\ll 1$ (here $\tau_{m}$ is the momentum relaxation
time and $\bar{p}$ is the characteristic momentum of carriers), leads only to
a weak anisotropy of emission. Such weak variations of polarization are determined
by the angle, $\phi$, with respect to the direction of the in-plane applied field,
and pertinent contributions are omitted below.

The paper is organized as follows. The basic equations governing the emission
of radiation in graphene are considered in Sec. II.
In Sec. III, we analyze the polarization characteristics and the spectral dependencies of radiation. The
concluding remarks and discussion of the assumptions used are given in the last
section.

\section{Basic equations}
Emission of radiation due to the interband transitions of nonequilibrium carriers
in graphene is described by the Wigner distribution function of photons,
$N_{\bf Qr}^{\mu\mu '}$ with the polarization indexes $\mu=1, 2$. Outside
of the graphene sheet, $N_{\bf Qr}^{\mu\mu '}$ obeys the following kinetic equation \cite{8}
\begin{equation}
{\bf v}_{\bf Q}\cdot\nabla_{\bf r}N_{\bf Qr}^{\mu\mu '}=J_R .
\label{1}
\end{equation}
Here ${\bf v}_{\bf Q}=\partial\omega_Q /\partial{\bf Q}=\widetilde{c}{\bf Q}/Q$
is the velocity of photon with the frequency $\omega_Q$, $\textbf{Q}\equiv (\textbf{q},
q_{\bot})$ is the wave vector of photon with the in-plane (transverse) component,
$\textbf{q}$ ($q_{\bot}$), and $\widetilde{c}$ is the speed of light in the medium
around the graphene sheet with dielectric permittivity $\epsilon$. The collision
integral $J_R$ describes the relaxation of photon distribution outside of graphene
(the glancing photons, with $q_{\bot} \to 0$, are not considered). The boundary
condition, around the graphene sheet placed at $z=0$, takes form
\begin{equation}
\upsilon_{\bot}N_{\bf Qr}^{\mu\mu '}\left|_{z=-0}^{z=+0}=I_{\bf Q}^{\mu\mu '} \right. ,
\label{2}
\end{equation}
where $\upsilon_{\bot}=\widetilde{c}q_{\bot}/Q$ and $I_{\bf Q}^{\mu\mu '}$ determines
the speed of spontaneous emission due to interband transitions. Within the
collisionless approximation, $I_{\bf Q}^{\mu\mu '}$ is determined by the direct
interband transitions ($l\neq l'$) between the states $|l{\bf p}\rangle$
and $|l'{\bf p}\rangle$, with distribution functions $f_{l{\bf p}}$ and $f_{l'{\bf p}}$,
and is given by \cite{8}
\begin{eqnarray}
I_{\bf Q}^{\mu \mu '}=4\frac{(2\pi e)^2}{\epsilon L^2\omega_Q}\sum
\limits_{ll'{\bf p}}\left\langle l{\bf p}\left|{\bf e}_{{\bf Q}\mu}
\cdot {\bf \hat v}\right|l'{\bf p}\right\rangle ^* \left\langle l{\bf p}
\left|{\bf e}_{{\bf Q}\mu '}\cdot{\bf \hat v}\right|l'{\bf p}\right\rangle
 \nonumber  \\
\times f_{l{\bf p}}(1-f_{l'{\bf p}})\delta (\hbar\omega_Q +\varepsilon_{l'p}
-\varepsilon_{lp} ) . ~~~~
\label{3}
\end{eqnarray}
Here the factor 4 comes from the spin and the valley degeneracy, $L^2$ is the
normalization area, and the polarization vectors, ${\bf e}_{{\bf Q}\mu}$,
are defined by relations $(\textbf{Q}\cdot \textbf{e}_{{\bf Q}\mu})=0$ and
$(\textbf{e}_{{\bf Q}\mu}^{\ast}\cdot \textbf{e}_{{\bf Q}\mu^{\prime}})=
\delta_{\mu\mu '}$. It is convenient to choose them with the angles
$\pm \pi/4$ towards the plane formed by the vectors $\textbf{Q}$ and $\textbf{n}$,
see Fig. 1. Point out that in Eq. (\ref{3}), due to the energy conservation,
contribute only the direct interband transitions between the states with $l=1$
and $l^{\prime}=-1$, where the $\delta$-function obtains the form
$\delta(\hbar \omega_{Q}-2v_{W}p)$.

Thus, the characteristics of emission (its intensity, spectral dependency, and polarization)
are expressed via $N_{\bf Qr}^{\mu\mu '}$ determined by Eqs. (\ref{1})-(\ref{3}).
The energy flow density is defined by the standard formula
\cite{8,12}
\begin{equation}
{\bf G}_{\bf r}=\sum\limits_\mu\int\frac{d{\bf Q}}{(2\pi )^3}{\bf v}_{\bf Q}
\hbar\omega _Q N_{\bf Qr}^{\mu\mu} .
\label{4}
\end{equation}
The polarization properties are characterized by the Stokes parameters,
$\xi_{x}$, $\xi_{y}$, and $\xi_{z}$ which are introduced by relations \cite{12}
\begin{eqnarray}
\xi_{x}=\frac{N_{\bf Q}^{12}+N_{\bf Q}^{12\ast}}{N_{\bf Q}} , \;\;\;
\xi_{y}=i\frac{N_{\bf Q}^{12}-N_{\bf Q}^{12\ast}}{N_{\bf Q}}, \nonumber \\
\xi_{z}=\frac{N_{\bf Q}^{11}-N_{\bf Q}^{22}}{N_{\bf Q}} , ~~~~~~~~~~~~~~~~
\label{5}
\end{eqnarray}
where $N_{\bf Q}=\sum_{\mu}N_{\bf Q}^{\mu\mu}$ determines an intensity of
radiation propagated along $\bf Q$.
Below we restrict ourselves by the nonabsorbing
medium, $J_R \to 0$, and consider (a) the in-plane homogeneous geometry, which is corresponded to a large-area sample of epitaxial graphene, and (b) a small-size sample
of exfoliated graphene.

For the case (a),  Eqs. (\ref{1}) and (\ref{2}) have the solution
\begin{equation}
N_{{\bf Q}z}^{\mu\mu^{\prime}}=
\frac{I_{\bf Q}^{\mu\mu '}}{\upsilon_\bot }
\left\{ \begin{array}{*{20}c}{\theta (z),} & {q_\bot >0}  \\
{\theta (-z),} & {q_\bot <0}  \\   \end{array} \right.
\label{6}
\end{equation}
which describes emitted radiation with $q_\bot >0$ ($q_\bot <0$) at $z>0$
($z<0$). The only nonvanishing component of the energy flow density $G_\bot$
is directed along $0Z$ and it is given by
\begin{equation}
G_\bot =\sum\limits_\mu\int\frac{d{\bf Q}}{(2\pi )^3}\hbar\omega_Q I_{\bf Q}^{\mu\mu}
=\int\limits_0^{\infty}d\omega\int\limits_{(q_\bot >0)} d\Omega_{\bf Q}\frac{\partial^2G}{\partial\omega\partial\Omega} ,
\label{7}
\end{equation}
where the differential flow, $\partial^2G/\partial\omega\partial\Omega$, is introduced
\begin{equation}
\frac{\partial^2G}{\partial\omega\partial\Omega}=\frac{\hbar\omega^3}
{(2\pi\tilde{c})^3}\sum\limits_\mu I_{\bf Q}^{\mu\mu} .
\label{8}
\end{equation}
In addition, it is convenient to introduce the frequency-dependent differential flow
$dG/d\omega =\int\limits_{(q_\bot >0)} d\Omega_{\bf Q}(\partial^2G/\partial\omega\partial
\Omega )$ that is averaged over the solid angle $\Omega_{\bf Q}$ subtended by the infinite plane.

Considering emission from the sample placed within the in-plane region $S_1$
[case (b)], we suppose that
$N_{\bf Q r}^{\mu\mu '}$ is zero outside of
$S_1$-region (i.e., we neglect by the edge diffraction effects). Thus, one obtains
the solution
\begin{eqnarray}
N_{\bf Qr}^{\mu\mu '}={\cal N}_{\bf Q}^{\mu\mu '}\left( x-\frac{q_x}{q_\bot}z,
y-\frac{q_y}{q_\bot}z\right), \\
{\cal N}_{\bf Q}^{\mu\mu '}(x,y)\equiv\frac{I_{\bf Q}^{\mu\mu '}}
{\upsilon_\bot} , ~~~~ (x,y) \in S_1 \nonumber
\label{9}
\end{eqnarray}
with ${\cal N}_{\bf Q}^{\mu\mu '}(x,y)=0$ outside of the $S_1$-region. In the
far zone, $R\gg\sqrt{S_1}$ (see Fig. 1b), the tangential components of $\bf G$
vanish and the radial component of energy flow takes form:
\begin{equation}
G_r(\theta )=\frac{S_1}{R^2}\int_0^{\infty}d\omega\frac{\partial^2G}{\partial\omega
\partial\Omega} ,
\label{10}
\end{equation}
where the differential flow, introduced by Eq. (\ref{8}), appears.

It is convenient to rewrite the speed of spontaneous emission Eq. (\ref{3}) by
making the replacement from the band quantum numbers $(l,l')$ to the electron-hole
representation, when the distributions $f_{l {\bf p}}$ are substituted for
electron and hole distributions as: $f_{1{\bf p}}\to f_{e{\bf p}}$ and
$(1-f_{-1{\bf p}})\to f_{h-{\bf p}}$. Then Eq. (\ref{3}) obtains the form
\begin{eqnarray}
I_{\bf Q}^{\mu\mu '}=4\frac{(2\pi ev_W )^2}{\epsilon L^2\omega_Q}\sum\limits_{\bf p}
M_{\mu\mu '}(\varphi ) ~~~~~ \nonumber \\
\times f_{e{\bf p}}f_{h,-{\bf p}}\delta (\hbar\omega_Q-2v_Wp) , ~~~~~~ \nonumber \\
M_{\mu\mu '}(\varphi )={\left\langle {1{\bf p}\left| {{\bf e}_{{\bf Q}\mu }
\cdot {\bf \hat v}} \right|-1{\bf p}}\right\rangle ^* \left\langle {1{\bf p}
\left| {{\bf e}_{{\bf Q}\mu '}\cdot{\bf \hat v}}\right|-1{\bf p}} \right\rangle } ,
\label{11}
\end{eqnarray}
where the interband matrix elements $M_{\mu\mu '}(\varphi )$ are dependent only on
the orientation of the unit vectors of polarization and on the angle $\varphi$,
giving the orientation of the momentum, ${\bf p}=(p\cos\varphi, p\sin\varphi )$.\cite{13}
Neglecting a weak anisotropy of distributions $f_{e,h{\bf p}}$, one can use in Eq.
(\ref{11}) the averaged over the $\bf p$-plane angle (such an average, over the angle $\varphi$,
is denoted using the overline) matrix element
\begin{equation}
\overline{M}_{\mu\mu '}=\frac{e_\mu^x e_{\mu '}^x+e_\mu^y e_{\mu '}^y}{2}
=\frac{1}{2}\left(\delta_{\mu\mu '}-\frac{\sin^2\theta}{2}\right) ,
\label{12}
\end{equation}
where $\theta=\widehat{{\bf n},{\bf Q}}$, see Fig. 1. As a result, the speed of
spontaneous emission, Eq. (\ref{11}),  we can rewrite as $I_{\omega\theta}^{\mu\mu '}
\equiv I_{\bf Q}^{\mu\mu '}$,
due to its dependence only on $\omega$ and $\theta$.
Finally, completing in Eq. (\ref{11}) the
integral over $p$ by using the energy
$\delta$-function, we obtain the speed of emission
as
\begin{equation}
I_{\omega\theta}^{\mu\mu '}=\frac{(2\pi ev_W )^2}{2\epsilon\omega}
\overline{M}_{\mu\mu '}f_{ep_\omega}f_{hp_\omega}\rho_{\hbar\omega /2} ,
\label{13}
\end{equation}
where $\rho_E=2E/\pi (\hbar v_W)^{2}$ is the density of states
and the distributions
are taken at the characteristic momentum $p_\omega$. Therefore, the spectral and polarization dependencies of radiation are presented
in Eq. (\ref{13}) by the
separate factors.

\section{Emission characheristics}
Here we study the polarization or the spectral characteristics of
the luminescence determined by Eqs. (\ref{5}) or (\ref{4}), (\ref{7}), and (\ref{9}), respectively. Distributions of nonequilibrium electrons and holes are described by the
quasiequilibrium Fermi functions $f_{kp}\approx\left\{\exp\left[\left( v_W p-
\mu_{k}\right)/T_{c}\right]+1\right\}^{-1}$ with effective temperature $T_{c}$
and the chemical potentials $\mu_{k}$, that determine the concentrations of
electrons and holes, $n_e$ and $n_h$. \cite{5} Instead of the concentrations $n_{e,h}$
it is convenient to introduce the
surface charge $e\Delta n=e(n_e-n_h)$, that
is defined by the gate voltage $V_g$, and the total concentration $n=n_e+n_h$,
that is defined by a character of
the generation-recombination processes. In
addition, the effective temperature $T_c$ of the hot electrons and holes can be
estimated from experimental data \cite{14} and from calculations.\cite{5,6}

\subsection{Polarization of radiation}
First, let us consider the polarization characteristics of radiation emitted.
Here we
 use that the spectral and polarization dependencies are separated, see
Eqs. (\ref{6}) and (\ref{13}), and the Stokes parameters (\ref{5}) are expressed
only through $\overline{M}_{\mu\mu '}$, Eq. (\ref{12}). Therefore
 the polarization
of radiation is independent of the frequency or the character of the
carriers
distributions. Then, for the geometry of Fig 1b, in the far-zone by using Eq.
(\ref{13}) in Eq. (\ref{5}) we obtain that $\xi_{y}=\xi_{z}=0$ and
\begin{equation}
\xi_x=\frac{2\overline{M}_{12}}{\overline{M}_{11}+\overline{M}_{22}}=
-\frac{\sin^2\theta}{1+\cos^2\theta} ,
 \label{14}
\end{equation}
which determines a degree of the linear polarization as function of $\theta$.
For the geometry of Fig. 1a for any plane wave contribution, with given ${\bf Q}$,
we again can introduce the Stokes parameters Eq. (\ref{5}) and they have the same
form as for the geometry of Fig 1b.

From Eq. (\ref{14}) it follows that for the normal propagation, $\theta\to 0$,
the emitted radiation becomes
nonpolarized, in agreement with the absence of any preferential direction over the graphene sheet. If to take into account a weak
lateral anisotropy of the carriers distributions induced by
the applied electric
field ${\bf E}$ then additional dependence of the polarization characteristics
appears from the mutual orientation of the vectors $\bf E$ and $\bf Q$ defined by
the angle $\phi$, see Fig. 1c.
This small addendum also depends on $\omega$ and
the form of the distribution functions of the carriers. For the glancing propagation,
at $\theta\to\pi /2$, Eq. (\ref{14}) shows that emitted radiation
is fully linearly polarized, in a 2D plane parallel to the graphene sheet. Appearance of the universal
angular dependence Eq. (\ref{14}) allows for separation of the
interband contributions
in graphene from any possible background emission (e.g., from a substrate,
a cover
layer or a gate).

\subsection{Spectral dependencies}
Next, we study the spectral dependencies and the efficiency of radiation emitted
by hot carriers for the geometries shown in Figs. 1a and 1b. Performing the
summation over polarization of Eq. (\ref{13}), we present  the differential flows
$d G/d \omega$ and $\partial^2G/\partial\omega\partial\Omega$, see Eq. (\ref{8}),
through $\sum_\mu I_{\omega\theta}^{\mu\mu}$. These differential flows give the
relevant intensities of radiation, by Eqs. (\ref{7}) and (\ref{10}). The differential
flow Eq. (\ref{8}) obtains the form
\begin{equation}
\frac{\partial^2 G}{\partial\omega\partial\Omega}=G_c A\left(\frac{\hbar
\omega}{2T_c}\right)\left( 1+\cos^2 \theta\right) ,
\label{15}
\end{equation}
where it is introduced the characteristic density of energy $G_c =\sqrt{\epsilon}
e^2(T_c/\hbar c)^3/\pi^2$ and the spectral behavior is determined by the following dimensionless function
\begin{equation}
A\left(\frac{\hbar\omega}{2T_c} \right) =\left(\frac{\hbar\omega}
{2T_c}\right)^3 f_{ep_\omega} f_{hp_\omega} .
\label{16}
\end{equation}
The angular dependence in Eq. (\ref{15}) is given by the factor $(1+\cos^2 \theta)$; i. e.,
the intensity of the glancing emission there is two times smaller than the normal
one. The differential flow Eq. (\ref{15}) integrated over the solid angle of half-space gives
\begin{equation}
\frac{dG}{d\omega}=\frac{8\pi}{3} G_c A\left(\frac{\hbar\omega}{2T_c}\right) ,
\label{17}
\end{equation}
i.e., $dG/d\omega$ is expressed through the function Eq. (\ref{16}).
For $T_c=$300 K and $\epsilon\simeq$3 we estimate the characteristic density
of energy, $G_c\propto T_c^3$, as $G_c\simeq 1.06\times 10^{-17}$ J/cm$^2$.
\begin{figure}[ht]
\begin{center}
\includegraphics{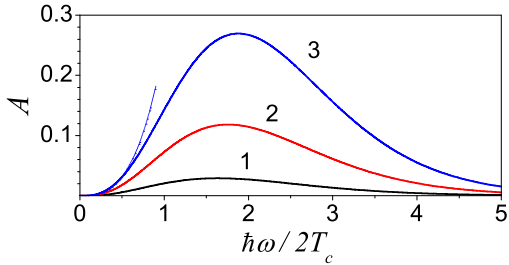}
\end{center}\addvspace{-1 cm}
\caption{(Color online) Dimensionless function $A$ versus $\hbar\omega /2T_c$
for the intrinsic graphene case at different generation-recombination levels
determined by ratio: $n/n_{T_c}=$0.5 (1), 1 (2) and 1.5 (3).}
\end{figure}

Thus, the spectral characteristics of radiation are expressed through the function
$A(\hbar\omega /2T_c)$ and it is shown in Fig. 2 for an intrinsic graphene. Here,
one defines the degree of nonequilibrium of the electron-hole pairs concentration
as the ratio, $n/n_{T_c}$, of the total concentration $n$ to pertinent equilibrium
one (at $T_c$ and $\mu_{e}=\mu_{h}=0$), where $n_{T_c}=(\pi/3)(T_{c}
/\hbar v_{W})^2$. This ratio also characterizes the effectiveness of generation-recombination processes.
 It is seen from Fig. 2 that as $n/n_{T_c}$ grows the intensity
of emission essentially increases
and the spectral maximum of radiation, localized at $\hbar\omega\sim T_c$, is slowly shifted.
For high frequencies $A(\hbar\omega /2T_c)$, Eq. (\ref{16}), is exponentially decreasing
as $\exp (-\hbar\omega /T_c)$ and at low frequencies this function grows $\propto\omega^3$
(the latter asymptotic dependence is shown in Fig. 2 by the dashed curve, for $n/n_{T_c}=1.5$).
\begin{figure}[ht]
\begin{center}
\includegraphics{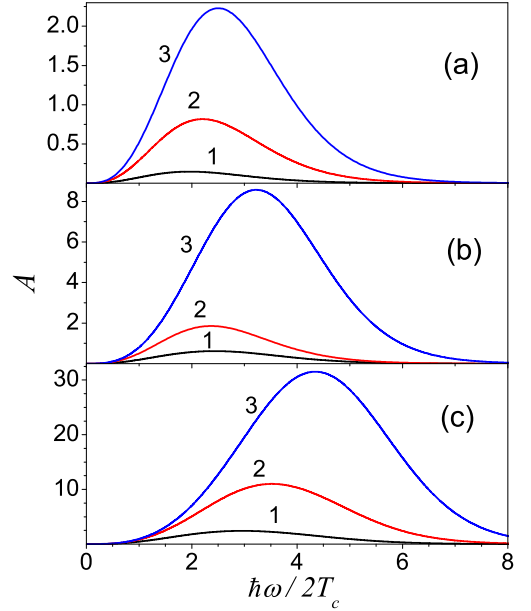}
\end{center}
\addvspace{-1 cm}
\caption{(Color online) Dimensionless function $A$ versus $\hbar\omega /2T_c$
for doped graphene at $T_c=$300 K and at different generation-recombination
conditions determined by ratio: $n/\Delta n=$1.25 (1), 2 (2) and 3 (3). Doping
levels are governed by gate voltages: $V_g=$5 V (a), 10 V (b), and 20 V (c).}
\end{figure}

For the doped graphene the spectral maximum is shifted to higher energies due
to
the Pauli blocking effect arising as the gate voltage, $V_g$, increases (compare
Figs. 3a, 3b, and 3c plotted for $T_c=$300 K). Here the total concentration of
carriers is defined by the ratio $n/\Delta n$, where $e\Delta n$ gives the
surface
 charge density controlled by $V_g$. Present calculations are conducted
for a typical
graphene structure on SiO$_2$ substrate of the thickness 300 nm.
As in intrinsic graphene, for growing concentration the intensity of emission
increases,
moreover, for growing $V_g$ the maximum also rapidly becomes larger.
For high energies
the spectrum of emission becomes exponentially decreasing.
\begin{figure}[ht]
\begin{center}
\includegraphics{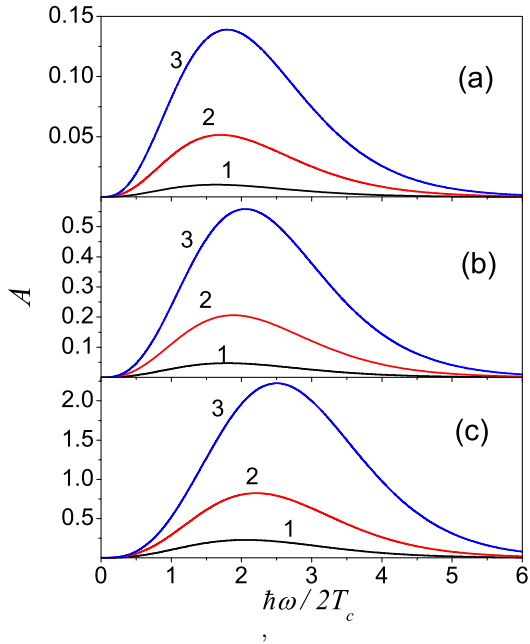}
\end{center}\addvspace{-1 cm},
\caption{(Color online) The same as in Fig. 3 at $T_c=$600 K}
\end{figure}

For increasing temperatures the character of dependences from $n/\Delta n$ and
$V_g$ is not modified, see Figs. 4a-c plotted for $T_c=$600 K. Here, for two times
larger temperature the altitude of $A$
becomes about one order of magnitude smaller. However, as $G_c\propto T_c^3$, the maximal differential flows Eqs. (\ref{15}) and
(\ref{17}) are weakely modified  for
such changes of $T_c$. In addition, point out
that here the position of maximum is more strongly
dependent on the character of
generation-recombination processes than for the intrinsic graphene
(compare  Fig. 2
with Figs. 3 and 4).

\subsection{Energy flow}
Now, let us consider the integral energy flow for the cases (a) and (b), when
Eqs. (\ref{7}) and (\ref{10}) can be expressed through the differential flow, Eq.
(\ref{15}). For the homogeneous case (a) the energy flow density takes form:
\begin{equation}
G_\bot =\frac{8\pi}{3}G_c\int\limits_0^\infty d\omega A\left(\frac{\hbar\omega}
{2T_c}\right) .
\label{18}
\end{equation}
For intrinsic graphene at $T_c=$300 K we calculate from Eq. (\ref{18}) that the flow $G_\bot\simeq$0.48 mW/cm$^2$, 2.02 mW/cm$^2$, and 4.69 mW/cm$^2$ corresponds,
respectively, to $n/n_T=$0.5, 1, and 1.5; these $n/n_T$ are used in Fig. 2. For
$T_c=$600 K and the same values of $n/n_T$, we obtain that the flow $G_\bot\simeq$
7.68 mW/cm$^2$, 32.3 mW/cm$^2$, and 75.0 mW/cm$^2$.
\begin{figure}[ht]
\begin{center}
\includegraphics{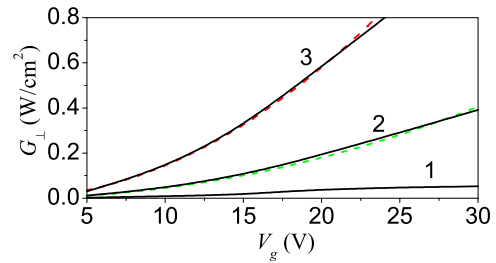}
\end{center}\addvspace{-1 cm}
\caption{(Color online) Energy flow density $G_\bot$ versus $V_g$ for temperature
$T_c=$300 K
at at different generation-recombination levels determined by
$n/\Delta n=$1.25 (1), 2 (2) and 3 (3). The dashed curves show the parabolic
approximation, $G_\bot\propto V_g^2$.}
\end{figure}

Fig. 5 shows that for doped graphene $G_\bot$ increases for growing concentrations of carriers,
defined by the gate voltage, whereas $n/\Delta n$ (or generation-recombination level) is fixed. From
Fig. 5 it is seen that for $V_g\sim$20 V, which corresponds to the concentration of carriers $>2 \times 10^{12}$ cm$^2$, the energy flow reaches the values $\sim 1$W/cm$^2$. In addition, the dependences $G_\bot$ versus $V_g$ are close to
parabolic ones (see the dashed curves in Fig. 5). The temperature dependence of $G_\bot$
is weak (less than 10\% for the $T_c$ increasing from 300 K to 600 K;
pertinent graphs
are not shown) because the spectral dependences are determined
by dimensionless parameter $\hbar\omega /T_c$.

For the geometry (b) the radial component of energy flow in the far-zone Eq. (\ref{10})
is expressed through $G_\bot$ as
\begin{equation}
G_r(\theta )=\frac{3S_1}{8\pi R^2}G_\bot\left( 1+\cos^2\theta\right) .
\label{19}
\end{equation}
Here the angular dependence coincides with that of Eq. (\ref{15}) and the energy flow
(\ref{19}) decreases $\propto R^{-2}$. Due to this, on macroscopic distances, for $\sqrt{S_1/R^2}\sim 10^{-3}$, it is possible
to register the energy flow $\sim 1 \mu$W/cm$^2$, while $G_\bot\sim 1$W/cm$^2$ is rather easily achievable according
to the above treatment.

\section{Discussion and concluding remarks}
Summazing the consideration performed, the examination of luminescence caused by
the interband transitions of hot electrons is presented. It is found, that the
universal frequency-independent polarization of emission is realized for a weakly anisotropic
distributions of nonequilibrium carriers. The spectral dependences of
radiation are determined by the factor $\propto\omega^3$ multiplied by the product
of electron and hole distributions.
These data, together with measurements of the
angular dependences and the integral intensity of radiation
(the latter strongly
depends on generation-recombination processes), allows the effective characterization
of hot carriers. In addition, obtained efficient emission of hot carriers
 in  mid-IR spectral regions opens a possibility for using the electroluminescence of graphene
as a source of radiation.

Now let us discuss recent experiments on emission by hot carriers from
the back-gated transistor structures, subjected to a strong in-plane electric field, \cite{14}
where measured spectral dependences are interpreted by using the Planck's law. For the latter
the high frequency asymptotic coincides with the asymptotic of Eq. (\ref{16}), for
the
case of the quasiequilibrium Fermi distributions. From these spectral dependences in
\cite{14} it was found the relation of $T_c$ with a power of the Joule heating. However,
the character of recombination, that could be determined from the intensity of radiation,
was not investigated;
dependences on a size of sample, angular dependences and
polarization characteristics
of the radiation also need a special treatment.

Next, we list and discuss the assumptions used. In the study of interaction of a
radiation with the
graphene sheet the only simplification used is the neglect by
attenuation of radiation propagated along the layer, so that obtained
results are
not applicable for $\theta\to\pi /2$. More rigid constraint is imposed by modeling
of the
 distributions of nonequilibrium carriers via the quasiequilibrium Fermi
functions, with given effective temperature and concentrations of carriers. This approximation shows that important information on the mechanisms
of energy relaxation
and recombination can be obtained from the spectra of luminescence. However,
for more
precise calculation of the spectral dependences more realistic distribution
 functions
of carriers will be needed. Further, the spectral and the polarization dependences
can be separated only in an approximation of a weak anisotropy of the distribution
of carriers.
 But such a separation can be broken under a strong enough electric field
when an essential anisotropy of the
distribution functions appears.

We also have
restricted our consideration by the homogeneous geometry (without taking into account
of the edge effects) and  the far-zone region geometry [approaches (a) and (b) in
Figs. 1a and 1b], however, the general treatment implies
solving of Eqs(\ref{1}),
(\ref{2}) for a specific geometry that is not well enough approximated by any of
these two limit geometries.

To conclude, obtained results show that spectral, angular, and polarization dependences
of the electroluminescence provide a convenient method of characterization of the hot
carriers in graphene (along with the electrooptical measurements
\cite{15} and a study
of the Raman scattering \cite{14,16}). Therefore present results will stimulate subsiquent
experiments and their theoretical interpretations designated for a verification of the relaxation mechanisms of
nonequilibrium carriers under their heating both for the
electric field and for the interband photoexcitation.


\end{document}